\documentclass[runningheads]{llncs}

\usepackage[T1]{fontenc}
\def\doi#1{\href{https://doi.org/\detokenize{#1}}{\url{https://doi.org/\detokenize{#1}}}}
\usepackage{graphicx}
\usepackage{hyperref}
\usepackage{listings}

\usepackage{verbatim}
\usepackage{setspace}
\usepackage{bbold}
\usepackage{tikz}
\def\checkmark{\tikz\fill[scale=0.4](0,.35) -- (.25,0) -- (1,.7) -- (.25,.15) -- cycle;} 

\begin{document}

\title{Supervised Contrastive Learning to Classify Paranasal Anomalies in the Maxillary Sinus}
%
%

\author{Debayan Bhattacharya \index{Bhattacharya, Debayan} \inst{1,2} , 
Benjamin Tobias Becker \index{Becker, Benjamin} \inst{2},
Finn Behrendt \index{Behrendt, Finn} \inst{1},
Marcel Bengs \index{Bengs, Marcel} \inst{1},
 Dirk Beyersdorff \index{Beyersdorff, Dirk}  \inst{3}, Dennis Eggert \index{Eggert, Dennis} \inst{2} ,
Elina Petersen \index{Petersen, Elina} \inst{4} ,
Florian Jansen \index{Jansen, Florian} \inst{2} ,
Marvin Petersen \index{Petersen, Marvin} \inst{5},
Bastian Cheng  \index{Cheng, Bastian} \inst{5} ,
Christian Betz \index{Betz, Christian} \inst{2} ,
Alexander Schlaefer* \index{Schlaefer, Alexander} \inst{1} ,
Anna Sophie Hoffmann* \index{Hoffmann, Anna}  \inst{2}  }
\index{Bhattacharya, Debayan}
\index{Becker, Benjamin}
\index{Behrendt, Finn}
\index{Bengs, Marcel}
\index{Beyersdorff, Dirk}
\index{Eggert, Dennis}
\index{Petersen, Elina}
\index{Jansen, Florian}
\index{Petersen, Marvin}
\index{Cheng, Bastian}
\index{Betz, Christian}
\index{Schlaefer, Alexander}
\index{Hoffmann, Anna}

\def\thefootnote{*}\footnotetext{These authors contributed equally to this work}
%
\authorrunning{D. Bhattacharya et al.}
%
\institute{Insititute of Medical Technologies and Intelligent Systems, Hamburg University of Technology, Hamburg, Germany \\
\email{debayan.bhattacharya@tuhh.de,d.bhattacharya@uke.de}\\
\and
Department of Otorhinolaryngology, Head and Neck Surgery and Oncology
\and
Clinic and Polyclinic for Diagnostic and Interventional Radiology and Nuclear Medicine
\and
Population Health Research Department, University Heart and Vascular Center
\and
Clinic and Polyclinic for Neurology \\
University Medical Center Hamburg-Eppendorf, Hamburg, Germany
}

\maketitle              
\begin{abstract}

Using deep learning techniques, anomalies in the paranasal sinus system can be detected automatically in MRI images and can be further analyzed and classified based on their volume, shape and other parameters like local contrast. However due to limited training data, traditional supervised learning methods often fail to generalize. Existing deep learning methods in paranasal anomaly classification have been used to diagnose at most one anomaly. In our work, we consider three anomalies. Specifically, we employ a 3D CNN to separate maxillary sinus volumes without anomaly from  maxillary sinus volumes with anomaly.  To learn robust representations from a small labelled dataset, we propose a novel learning paradigm that combines contrastive loss and cross-entropy loss. Particularly, we use a supervised contrastive loss that encourages embeddings of maxillary sinus volumes with and without anomaly to form two distinct clusters while the cross-entropy loss encourages the 3D CNN to maintain its discriminative ability. We report that optimising with both losses is advantageous over optimising with only one loss. We also find that our training strategy leads to label efficiency. With our method, a 3D CNN classifier achieves an AUROC of 0.85 \(\pm\) 0.03 while a 3D CNN classifier optimised with cross-entropy loss achieves an AUROC of 0.66 \(\pm\) 0.1. Our source code is available at \url{https://github.com/dawnofthedebayan/SupConCE_MICCAI_22}.

\end{abstract}
\section{Introduction}

Paranasal sinus anamolies are common incidental findings reported in patients who undergo diagnostic imaging of the head \cite{Wilson.2017} for neuroradiological assessment. Understanding the different opacifications of the paransal sinuses is very useful, because the frequency of these findings represent clinical challenges \cite{Hansen.2014}, and little is known about the incidence and significance of these morphological changes in the general population. 
There have been numerous studies on analysing the occurrence and progression of these incidental
findings\\ \cite{Tarp.2000,Rak.1991,Stenner.2014,Rege.2012,Cooke.1991}, but mostly in patients with sinunasal symptoms. 

In our study, elderly people (45 - 74 years) received an MRI for neuro-radiological assessment \cite{Jagodzinski2020} in the city of Hamburg.  The purpose of our study is to find out what percentage of patients, who do not have sinunasal symptoms, show findings in the paranasal sinus system in MRI images and if it is possible to detect anomalies in the paranasal sinus system using deep learning techniques and further analyze and classify based on their volume, shape and other parameters like local contrast.

A three year retrospective study showed malignant tumors were misdiagnosed as nasal polyps with a misdiagnosis rate of 5.63\% while inverted papilloma were misdiagnosed as nasal polyps with a rate of 8.45\% \cite{Ma.2012}.  As a first step, it would be beneficial to separate MRI with any paranasal anomaly from normal MRI using Computer Aided Diagnostics (CAD). This would allow the physicians to closely inspect the MRIs containing paranasal anomaly with finer detail. This can reduce chances of misdiagnosis while decreasing the workload of physicians from having to see normal MRIs.

There have been works of paranasal sinus anomaly diagnosis using supervised learning \cite{Kim.2019,Jeon.2021,Liu.2022}. However, all these works have been used to classify at most one anomaly. Our work considers three anomalies namely: (i) mucosal thickening (ii) polyps and (iii) cysts. The anomalies are differently located within the maxillary sinus and therefore we use a 3D volume of the maxillary sinus as input to our 3D CNN. Additionally, existing works train on large datasets to achieve good performance. However, labelling is a time consuming task and in some scenarios it requires the supervision of Ears, Nose and Throat (ENT) specialised radiologists who may not be immediately available. In our case, we have a small dataset of 199 MRI volumes. To mitigate the limitations of our small labelled dataset, we use contrastive learning \cite{DBLP:journals/corr/abs-1807-03748} to learn robust representations that does not overfit on the training set.  
 In contrastive learning, an encoder learns to map positive pairs close together in the embedding space while pushing away negative pairs. In SimCLR \cite{Chen.2020}, an image is transformed twice through random transformations. The transformed "views" of the image constitute a positive pair and every other image in the mini-batch is used to construct negative pairs with the reference image. There has been significant research in finding the best transformations as the chosen transformations dictate the quality of learnt representation \cite{Chen.2020}. The underlying assumption is that transformations augment the image while preserving the semantic information. However, in our case two or more images in the mini-batch can be semantically similar as they belong to the same class. Particularly, a maxillary sinus of one patient can be semantically similar to another patient's maxillary sinus if both patients do not exhibit any paranasal anomalies. Furthermore, they can be different as well due to the anatomical variations of the maxillary sinus \cite{Tarp.2000,Rege.2012}. Since our dataset contains anomalies, it is also important that the classifier does not overfit to a particular anomaly. Therefore, it is important for a classifier to learn anatomically invariant representation of the maxillary sinus for volumes that contain no anomaly and learn anomaly invariant representation of the maxillary sinus for volumes exhibiting one of the three anomalies. This can reduce the chances of overfitting on the training set. This motivates us to employ a supervised contrastive learning approach \cite{Khosla.2020} that brings embeddings of maxillary sinus volumes without anomaly closer together while pushing away embeddings of maxillary sinus with anomaly in the embedding space and vice versa. Compared to the original method \cite{Khosla.2020}, we propose a training strategy that simultaneously trains our 3D CNN using the supervised contrastive loss and regular cross-entropy loss using two different projection networks. The reasoning behind this is that minimizing the contrastive loss encourages the 3D CNN to learn representations that are robust to anatomical and anomaly variations while the cross-entropy enforces the 3D CNN to preserve discriminative ability.

 In summary, our contributions are three-fold. First, we demonstrate the feasibility of a deep learning approach to classify between normal and anomalous maxillary sinuses. Second, we demonstrate through extensive experiments that combining supervised contrastive loss and cross-entropy loss is the better approach to improve the discriminative ability of the 3D CNN classifier. Third, we empirically show that our method is the most label efficient.

\section{Method}
\begin{figure}[!htb]
\centering
\includegraphics[width=1.0\textwidth]{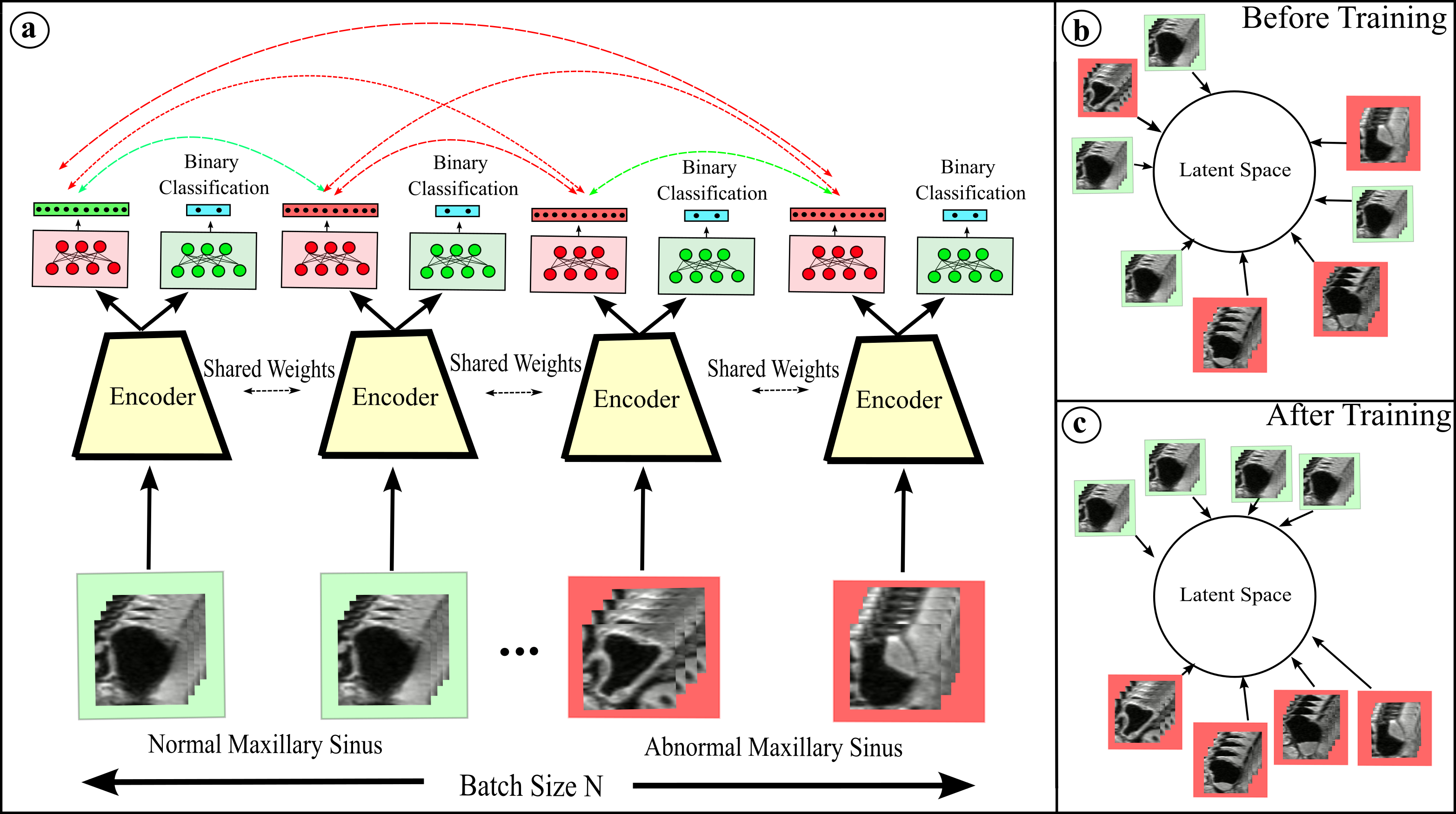}
\caption{(a) Our proposed method. The curved green and red lines represent similar and dissimilar representations respectively. (b)-(c) Illustration of the latent space embedding of normal and anomalous maxillary sinuses before and after the encoder is trained respectively.} \label{fig1}
\end{figure}

In global contrastive learning methods such as SimCLR \cite{Chen.2020}, we learn a parametric function \(F_{\theta} : X \rightarrow  \mathbb{R}^{D} \) where \(\mathbb{R}^{D}\) is a unit hypersphere. \(F_{\theta}\) is trained to map semantically similar samples closer together and semantically dissimilar samples further apart through the InfoNCE loss \cite{vanOord.2018}. The underlying assumption here is that images in the mini-batch are semantically dissimilar. This is untrue in our case as we have maxillary sinus volumes belonging to one of the two classes and there are semantic similarities and dissimilarities in the intra and inter class samples. Therefore, it is not possible to form meaningful clusters from the global contrastive loss described in SimCLR \cite{Chen.2020}. Therefore, to incorporate the class priors, we sample volumes from the two classes explicitly. Given an input mini-batch \(B = \{x_{1},x_{2},...,x_{N}\}\) where \(x_{i}\) represent the input 3D volume, a random transformation set \(T\) is used to form a pair of augmented volumes. Instead of randomly sampling \(N\) samples, we randomly sample \(N/2\) maxillary sinus volumes without anomaly  and \(N/2\) maxillary sinus volumes with anomaly from our dataset. Each of these \(N/2\) subsets undergo random transformation twice using \(T\). Let us denote the set containing all the augmented volumes of the \(C\) classes as \(\mathbb{M} = \bigcup_{c=1}^{C} M_{c} \). In our case, \(C = 2\). Here, \(M_{c}\) is the subset of augmented volumes belonging to a single class and \(|M_{c}|\) = \(N\) is its cardinality. Let \(m_{i}, i \in \mathbb{I} = \{1,2,...,2N\}\) represent the augmented volumes in set \(\mathbb{M}\) and \(m_{k(i)}\) is its corresponding volume augmented from the same volume in \(B\). Furthermore, let \(I_{c}\) represent indices of all the augmented volumes belonging to class \(c\) such that \(\mathbb{I} = \bigcup_{c=1}^{C} I_{c}\). Using the above stated assumptions, the InfoNCE loss that takes into consideration the class priors when making positive and negative pairs can be written as: 


\begin{equation}
    L_{simclr} = - \sum_{c=1}^{C} \frac{1}{|M_{c}|} \sum_{i \in I_{c}} log \frac{e^{sim(Z_{i},Z_{k(i)})/\tau}}{e^{sim(Z_{i},Z_{k(i)})/\tau} + \sum_{j \in \mathbb{I} \setminus I_{c}} e^{sim(Z_{i},Z_{j})/\tau} }
    \label{eq:2}
\end{equation}
where \(\tau \in \mathbb{R}^{+}\) is the scalar temperature parameter, \(Z_{i} = F_{\theta}^{con}(m_{i})\) is the normalised feature vector such that \(F_{\theta}^{con}(.) = Proj_{1}(Enc(.))\),   \(k(i)\) is the index of the corresponding volume in \(\mathbb{M}\) augmented from the same volume in \(B\) and \(sim(.)\) is the cosine similarity function.
Although Eq. \ref{eq:2} only constructs negative pairs such that the two volumes are from different classes, it still constructs positive pairs by augmenting the same volume using the random augmentation set \(T\). As a result, we are reliant on the transformations to learn meaningful representations. However, the transformations do not guarantee anatomical and anomaly invariance as the encoder is not incentivised to produce similar representations for volumes belonging to the same class in the mini-batch. Therefore, we use a supervised contrastive loss \cite{Khosla.2020} that constructs arbitrary number of positive pairs where each volume in the pair is from the same class but unique in the mini-batch. Formally, the supervised contrastive loss can be described as shown below: 

\begin{equation}
    L_{sc} = - \sum_{c=1}^{C} \frac{1}{|M_{c}|} \sum_{i \in I_{c}} log \frac{\sum_{j \in I_{c} \setminus \{i\}} e^{sim(Z_{i},Z_{j})/\tau}}{\sum_{j \in I_{c} \setminus \{i\}} e^{sim(Z_{i},Z_{j})/\tau} + \sum_{j \in \mathbb{I} \setminus I_{c}} e^{sim(Z_{i},Z_{j})/\tau} }
    \label{eq:3}
\end{equation}

The main differences of eq. \ref{eq:3} compared to eq. \ref{eq:2} is that numerous positive pairs are constructed in the numerator by matching every volume with every other volume belonging to the same class in the mini-batch. In this case, \(|M_{c}|\) = \(N/2\) as we do not use \(T\) and \(\mathbb{I} = \{1,2,...,N\}\). This incentivises the encoder to give similar representations for volumes belonging to the same class. This leads to learning anatomical and anomaly invariant representations.  Apart from using \(L_{sc}\) we also use regular cross-entropy loss to preserve the discrimintative ability of our 3D CNN. The cross-entropy loss is formalised as follows: 

\begin{equation}
    L_{ce} = -\frac{1}{N} \sum_{i \in \mathbb{I}}  y_{i}log(F_{\theta}^{class}(m_{i})) 
    \label{eq:4}
\end{equation} 
Here, \(F_{\theta}^{class}(.)\) can be decomposed into \(Proj_{2}(Enc(.))\) and \(y_{i}\) is the class label of \(m_{i}\) such that \(y_{i} \in \{0,1\}\). Therefore, our combined loss function is 
\begin{equation}
    L_{ours} = L_{sc} + \lambda L_{ce}
    \label{eq:5}
\end{equation} 

In our case, we set \(\lambda = 1\). In summary, we train models using only \(L_{ce}\) and set this as the baseline. We then train models using \(L_{simclr}\) to show that transformation invariance does not help in our downstream classification task. Next, we train our models using \(L_{sc}\) to show the benefit of clustering based on class priors and the importance of anatomical and anomaly invariant representation learning. Finally, we train our models using \(L_{ours}\) to show that contrastive loss and cross-entropy loss improve the discriminative ability of the models and overfit the least on the training set.

\subsection{Dataset}  

\begin{figure}[!htb]
\centering
\includegraphics[width=1.0\textwidth]{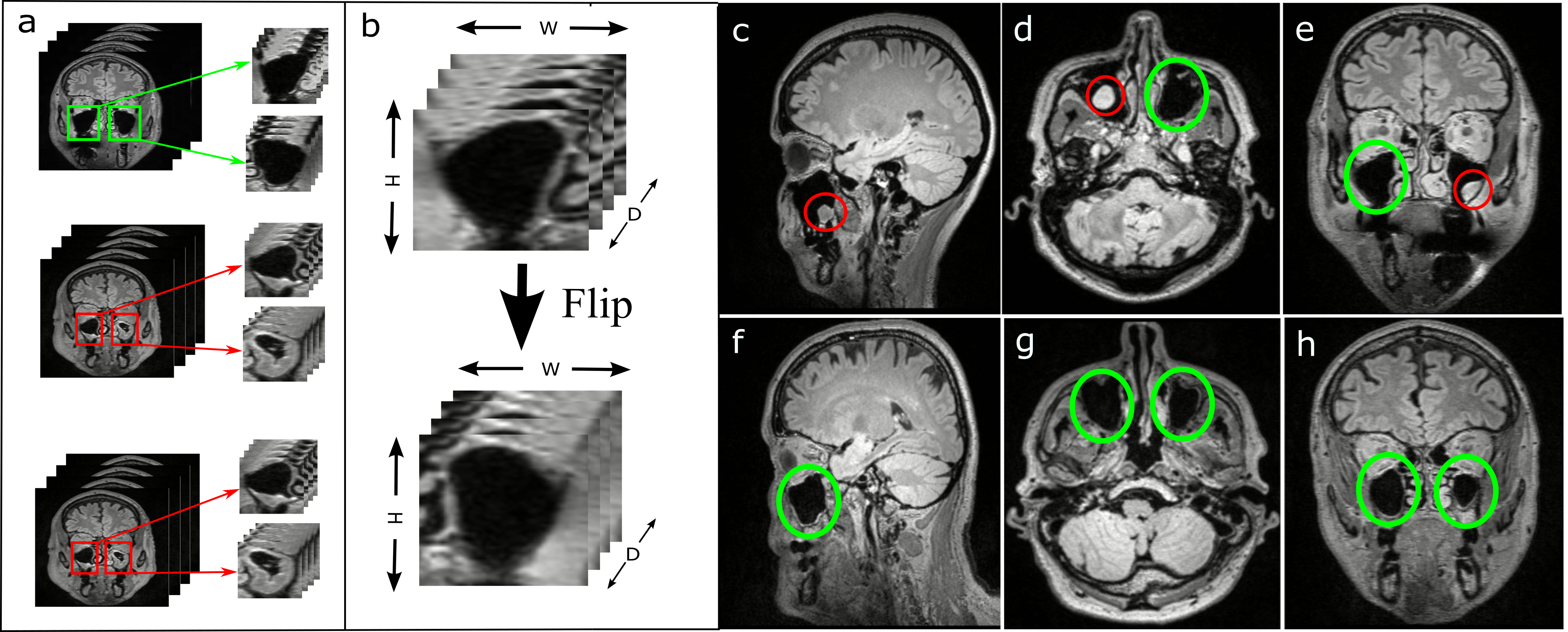}
\caption{[LEFT] Pre-processing steps involve (a) Extraction of 3D sub-volumes of left and right maxillary sinus from FLAIR MRI samples. (b) Flipping the coronal plane slices of right maxillary sinus sub-volume to give it the appearance of left maxillary sinus sub-volume.  [RIGHT] FLAIR-MRI slices from the sagittal , axial and coronal views illustrating the difference between anomaly and normal class. The red circles denote anomalies and green circles denote normal maxillary sinus. (c) Cyst observed in right maxillary sinus. (d) Polyp observed in the left maxillary sinus (e) Cyst observed in left maxillary sinus. (f) -(h) FLAIR-MRI slices with no pathology.} \label{fig2}
\end{figure}

As part of the population study \cite{Jagodzinski2020}, MRI of the head and neck area of participants based in the city of Hamburg, Germany were recorded. The MRIs were recorded at University Medical Center Hamburg-Eppendorf. The age group of the participants were between 45 and 74 years. Each participant had T1-weighted and fluid attenuated inversion recovery (FLAIR) sequences stored in the NIfTI\footnote{https://nifti.nimh.nih.gov/} format. FLAIR-MRIs were chosen as the imaging modality as the incidental findings are more visible due to the higher contrast relative to T1 weighted MRI. 
The labelled dataset consists of 199 FLAIR-MRI volumes of which 106 patients exhibit normal maxillary sinuses and 93 patients have maxillary sinuses with anomaly in at least one maxillary sinus. The diagnosis of the observed pathology in 199 FLAIR-MRIs was confirmed by two ENT surgeons and one ENT specialised radiologist. The incidental findings are categorised and defined as follows: (i) mucosal thickening (ii) polyps (iii) cysts. The statistics of the pathology observed is reported in the supplementary material. In this work, all the anomalies are grouped into a single class called "anomaly" and all the normal maxillary sinuses are grouped into a class called "normal". Altogether, there are 269 maxillary sinus volumes without anomaly and 130 abnormal maxillary sinus volumes with anomaly.  Each MRI has a resolution of 173 x 319 x 319 voxels along the saggital, coronal and axial directions respectively. The voxel size is 0.53 \(mm\) x 0.75 \(mm\) x 0.75 \(mm\). 

\textbf{Preprocessing}: We performed rigid registration by randomly selecting one FLAIR-MRI sample as a fixed volume followed by resampling to a dimension of 128 x 128 x 128. Of the resampled volumes, we extracted two sub-volumes, one for each maxillary sinus from a single patient. We made sure that the extracted sub-volumes subsumed the maxillary sinus. 

Since the maxillary sinus are symmetric, we horizontally flipped the coronal planes of right maxillary sinus volumes to make it look like left maxillary sinus volumes. The decision of which maxillary sinus to flip was arbitrary. Ultimately, these sub-volumes were reshaped to a standard size of 32 x 32 x 32 voxels for the 3D CNN. Finally, all the maxillary sinus volumes were normalised to the range of -1 to 1. Our preprocessing step is shown in Fig. \ref{fig2}. 
 
\textbf{Training, Validation and Test Split}: We perform a  nested stratified K-fold with 5 inner and 5 outer folds. The inner fold was used to choose the best hyperparameters. In summary, each experiment has 80 volumes in test set, 64 volumes in cross validation set and 255 samples in training set. The folds are constructed by preserving the percentage of samples for the two classes.

\section{Experiments and Discussion}

\subsection{Implementation Details}
All of our experiments are implemented in PyTorch \cite{Paszke.03.12.2019} and PyTorch Lightning \cite{falcon2019pytorch}. We use a batch size of 128 for all our experiments based on hyperparameter tuning (See supplementary material). We use Adam Optimization with a learning rate of 1e-4. Similar to Chen et al. \cite{Chen.2020}, we fix \(\tau=0.1\). Our encoder \(Enc(.)\) is the 3DResNet18 \cite{Hara.25.08.2017}. Our projection layer \(Proj_{1}(.)\) is a fully connected layer with input dimension 512 and output dimension 128. A ReLU activation is placed in between the layers. Projection layer \(Proj_{2}(.)\) is a linear fully connected layer with input and output dimensions of 512 and 2 respectively. All our models are trained for 200 epochs. 

\subsection{Evaluation of learnt representations}

\begin{table}[hbt!]
\centering

\caption{Evaluation of our representations}\label{tab1}
\resizebox{\columnwidth}{!}{%
\begin{tabular}{|c|c|c|c|c|c|c|c|}
\hline
Method & $L_{ce}$  & $L_{simclr}$  & $L_{sc}$ & Accuracy & F1(weighted) & AUROC & AUPRC\\
\hline
Scratch & \checkmark &   &   & $0.68\pm0.03$ & $0.57\pm0.04$ & $0.66\pm0.10$ & $0.53\pm0.12$ \\
\hline
Scratch(with aug) & \checkmark &    & & $0.69\pm0.03$ & $0.58\pm0.05$ & $0.68\pm0.10$ & $0.56\pm0.10$\\
\hline
SimCLR & &  \checkmark  & & $0.65\pm0.03$ & $0.58\pm0.04$ & $0.54\pm0.09$ & $0.41\pm0.09$\\
\hline
SupCon \cite{Khosla.2020} & &    & \checkmark &  $0.72\pm0.05$ & $0.70\pm0.06$ & $0.73\pm0.08$ & $0.61\pm0.08$ \\
\hline
Ours & \checkmark &    & \checkmark & $\textbf{0.80}\pm\textbf{0.02}$ & $\textbf{0.78}\pm\textbf{0.03}$ & $\textbf{0.85}\pm\textbf{0.03}$ & $\textbf{0.78}\pm\textbf{0.03}$\\
\hline

\end{tabular}
}
\end{table}
The metrics we have used to evaluate our representations are accuracy, F1 weighted, Area Under Receiver Operator Characteristics (AUROC) and Area Under Precision Recall Curve (AUPRC). We chose F1 weighted and AUPRC as they give a fair assessment of the performance of the models in the presence of class imbalance. Our random transformation set \(T\) consists of random affine, flip and gaussian noise as these are semantic preserving transforms. For training models with \(L_{simclr}\), \(L_{sc}\) and \(L_{ours}\) we followed the training strategy followed by Khosla et al \cite{Khosla.2020}. The inference is performed using \(Proj_{2}(.)\) for all the experiments. We test for statistically significant difference in our performance metrics using a permutation
test with \(nP\)= 10000 samples and a significance level of \(\alpha\) = 0.05 \cite{Efron.1998}. The difference in the AUROC, AUPRC, F1 weighted and accuracy of our method is significant (p < 0.05) compared to the other methods. From the results in Table \ref{tab1} we observe that the models trained with only \(L_{ce}\) overfit and do not generalize well. Models trained with \(L_{sc}\) achieve a significant boost in all metrics compared to models trained using \(L_{ce}\) and \(L_{simclr}\). We conjecture this to be the case because the supervised contrastive loss clusters the maxillary sinus volumes representations based on its class leading to invariant representations and less overfitting. The absence of \(L_{ce}\) causes the clusters to be more spread out (See Fig. \ref{fig3} SupCon). Our loss \(L_{ours}\) shows the best performance due to the increased discriminative ability of the classifier which is reflected by the formation of compact clusters (See Fig. \ref{fig3} Ours). SimCLR performs the worst as they fail to form meaningful clusters (See Fig. \ref{fig3} SimCLR).

\subsection{Label Efficient Representation Learning}

\begin{figure}[hbt!]
\centering
\includegraphics[width=1.0\textwidth]{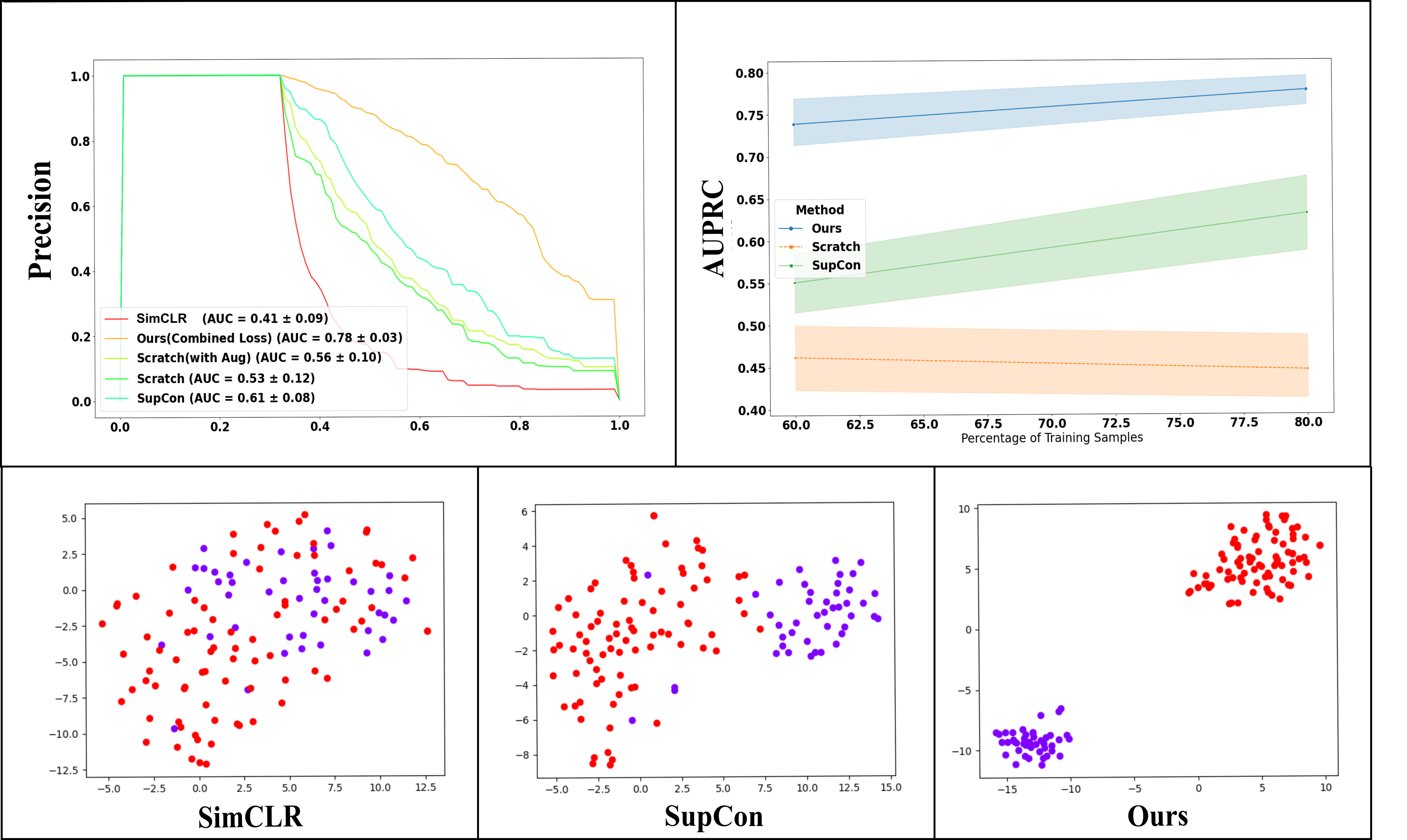}
\caption{(UPPER LEFT) Precision Recall Curve. (UPPER RIGHT) Illustrating the label efficiency of our method by plotting the AUPRC against the percentage of training dataset (BOTTOM LEFT, MIDDLE, RIGHT) t-SNE \cite{JMLR:v9:vandermaaten08a} with perplexity = 30, learning rate = 200, iterations=1000 used to visualise the representations learnt by various contrastive losses. The red dots denote normal class and purple dots denote anomaly class.} \label{fig3}
\end{figure}
We evaluated the performance of models trained with the different loss functions by supplying 60\% and 80\% of training samples. We excluded SimCLR  approach because it performed very poorly on 60\% and 80\% of training set. We performed a five-fold cross validation experiment with the same training strategy. The test set is the same for all the experiments. The AUPRC of the models are displayed in the upper right graphic of fig. \ref{fig3}. We observe that models trained using \(L_{sc}\) outperform the baseline by a significant margin. Even with limited data, both the losses ( \(L_{sc}\) and  \(L_{ours}\)) show improved performance with the injection of more training data. We also observe that \(L_{ours}\) almost achieves AUPRC of 100\% training set and it is also the most label-efficient. These results reveal that our approach can reduce labelling effort of physicians to an extent thereby allowing them to invest more time in the diagnosis and evaluation of difficult clinical cases.

\section{Conclusion}

Previous works on population studies \cite{Tarp.2000,Rak.1991,Stenner.2014,Rege.2012,Cooke.1991} have relied on manual analysis by physicians. Our work is a first step towards bringing automation in such studies. We show the benefit of contrastive loss in the classification of anomalies in the maxillary sinus from limited labelled dataset. Furthermore, we report the performance improvements relative to regular cross-entropy. Specifically, clustering based on class priors is helpful to learn representations that overfit less on the training set. We also show that a combinination of cross-entropy loss and supervised contrative loss improve the performance of the model. Finally, compared to other works that use deep learning for paranasal inflammation study \cite{Jeon.2021,Kim.2019,Liu.2022}, ours is the first work that tries to achieve label efficiency and thus attempts to reduce the workload of physicians. A limitation of our work is that the classification accuracy is still not satisfactory. As future work, we plan to label more MRIs and even perform classification of the type of anomaly observed in the maxillary sinus.

%
%
%
\bibliographystyle{splncs04}
\bibliography{bibliography}

\end{document}